# Numerical solution of moving plate problem with uncertain parameters

S. Nayak and S. Chakraverty[*]
Department of Mathematics, National Institute of Technology Rourkela, Odisha -769008, India

**Abstract**

This paper deals with uncertain parabolic fluid flow problem where the uncertainty occurs due to the initial conditions and parameters involved in the system. Uncertain values are considered as fuzzy and these are handled through a recently developed method. Here the concepts of fuzzy numbers are combined with Finite Difference Method (FDM) and then Fuzzy Finite Difference Method (FFDM) has been proposed. The proposed FFDM has been used to solve the fluid flow problem bounded by two parallel plates. Finally sensitivity of the fuzzy parameters has also been analysed.

1. Introduction

Uncertainty is a common phenomenon which arises in various branches of engineering and science. These are vague, imprecise and incomplete information about the variables and parameters being a result of errors in measurement, observations, experiment, applying different operating conditions or it may be maintenance induced errors etc. These uncertainties and vagueness are traditionally handled by stochastic approaches where the parameters and variables are considered as random. As a result, the governing differential equations involve randomness and turn to be stochastic.

[*] Corresponding author
E-mail address: sne_chak@yahoo.com (S. Chakraverty), sukantgacr@gmail.com (S. Nayak)

In this context, various investigations have been done by considering stochastic nature of the involved parameters. Researchers have used both the exact and numerical approaches to solve different stochastic problems. In the following paragraphs, the numerical approach to solve stochastic differential equations are discussed which are closely related to the present work. Numerical methods viz. finite difference, finite element and perturbation methods in connection with stochastic variables are investigated by various authors. Ghanem and Spanos [1] described a natural extension of the deterministic finite element method. They considered spectral approach to determine the system response. This solution approach involves two stages. The first stage consists of adequately representing the stochastic process corresponding to the random system properties. The second stage involves solving of the resulting set of equations. Hien and Kleiber [2] used variation principle with the stochastic method to solve stochastic differential equations for transient and steady-state heat problems. Kaminski [3] considered second order perturbation method to handle random parameters stochastically to solve structural related problems. In [4, 5], various heat transfer problems have been investigated within uncertain environment. Xiu [4] presented a random spectral decomposition method for the solution of transient heat conduction subjected to random inputs. Emery [5] combined boundary element method with orthogonal expansion theory and proposed a new numerical method for solving stochastic heat transfer problems. Further, Kaminski and Carey [6] solved fluid flow problem by considering traditional second order stochastic perturbation technique.

In the above literature review, it has been observed that we need large number of experimental data to investigate problem stochastically. In practice, it may be difficult sometime to get a large number of experimental data so we need an alternative method in which we may handle the uncertainty with few experimental data. In this regards, Zadeh [7] proposed an alternate idea i.e. fuzzy number to handle uncertain values which helps to overcome the complexity of availability of large data. Hence we need the help of interval/fuzzy analysis for managing these types of data. The direct implementation of interval or fuzzy becomes complex and the computation is also a difficult task. So to avoid such difficulty various authors tried different techniques to handle such difficulty. In this context, Dong and Shah [8] proposed vertex method for computing functions of fuzzy variables. In vertex method, various possible combinations of uncertain values are considered and examined. For *n* intervals it need to perform $2^n$ computations and then the minimum and maximum values are assigned as the left and right bounds respectively. Whereas, Dong and Wong [9] used Fuzzy Weighted Average Method (FWAM) to handle uncertain fuzzy variables. Further, Yang et al. [10] modified FWAM and proposed methods which require less computation. Klir [11] revised fuzzy arithmetic by considering the relevant requisite constraints and a transformation method based on the concept of $\alpha$-cut has been presented by Hanss [12]. This transformation method reduces fuzzy values into interval form and then used.

Further, it may be reveals from the above literature that interval/fuzzy arithmetic is a great tool to handle the uncertainty. This arithmetic have been used by various authors to investigate different engineering problems and few of these are discussed here. Bart et al. [13] investigated the uncertain solution of heat conduction problem and gave a good comparison between response surface method and other methods. Further, various diffusion problems are investigated in [14, 15] using proposed fuzzy finite element method. Next, Wang and Qiu [16] considered fuzzy finite difference method to solve uncertain heat conduction problems.

It has also been seen that the insertion of uncertainty makes the problem complex to handle, so to avoid such complexity various authors proposed alternate techniques which are sometimes a tedious task to perform. In view of this, here a non-probabilistic finite difference method viz. fuzzy finite difference method is proposed. This method converts the fuzzy parameters into intervals and then these interval values are transformed into crisp form using the proposed transformation in [17]. In addition, the stability of the proposed method has been analysed. Using this method a moving plate problem is investigated and uncertain solutions are depicted graphically. Finally, present results are validated through comparing with the crisp results and it is found in good agreement. Along with the validation, the variability of uncertainty has also been studied in various cases.

In this paper, the basic definition of fuzzy number and its arithmetic operations have been presented in section 2. Section 3 describes the statement of the problem along with the initial and boundary conditions. In section 4, the connection of uncertainty with the well-known finite difference method has been discussed. Then the stability of the proposed method is analyzed in section 5. The proposed method is demonstrated for the said problem and the sensitivity of uncertain parameters are described as various cases in section 6.

## 2. Preliminary

The basic definitions of a fuzzy number are given in [18-20] as follows

Definition 2.1

A fuzzy number $x$ in parametric form is a pair $[\underline{x}, \bar{x}]$ of functions $\underline{x}(\alpha)$ and $\bar{x}(\alpha)$, where $\alpha \in [0, 1]$ satisfies the following conditions.

i. $\underline{x}(\alpha)$ is a bounded non-decreasing left continuous function in (0, 1], and right continuous at 0.

ii. $\bar{x}(\alpha)$ is a bounded non-increasing left continuous function in (0, 1], and right continuous at 0.

iii. $\underline{x}(\alpha) \leq \bar{x}(\alpha)$, $0 \leq \alpha \leq 1$.

Definition 2.2

A fuzzy number $\tilde{A} = [a^L, a^N, a^R]$ is said to be triangular fuzzy number when the membership function is given by (Fig. 1)

$$\mu_{\tilde{A}}(x) = \begin{cases} 0, & x \leq a^L; \\ \dfrac{x - a^L}{a^N - a^L}, & a^L \leq x \leq a^N; \\ \dfrac{a^R - x}{a^R - a^N}, & a^N \leq x \leq a^R; \\ 0, & x \geq a^R. \end{cases}$$

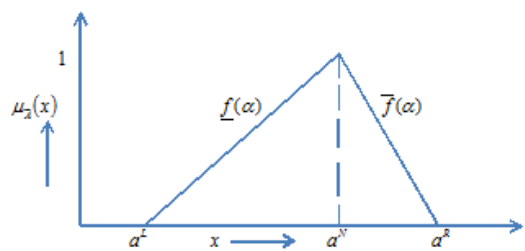

Fig 1. Triangular Fuzzy Number (TFN)

Triangular fuzzy number $\tilde{A} = [a^L, a^N, a^R]$ may be transformed into interval form by using $\alpha-$cut as follows

$$\tilde{A} = [a^L, a^N, a^R]$$
$$= [a^L + (a^N - a^L)\alpha, a^R - (a^R - a^N)\alpha], \quad \alpha \in [0, 1];$$

Definition 2.4

If the fuzzy numbers are taken in ordered pair form (as discussed in definition 2.1) then using limit method [21], the arithmetic rules may be written as

1. $[\underline{x}(\alpha), \bar{x}(\alpha)] + [\underline{y}(\alpha), \bar{y}(\alpha)]$

    $= [\min \{\lim_{n\to\infty} m_1 + \lim_{n\to\infty} m_2, \lim_{n\to 1} m_1 + \lim_{n\to 1} m_2\}, \max \{\lim_{n\to\infty} m_1 + \lim_{n\to\infty} m_2, \lim_{n\to 1} m_1 + \lim_{n\to 1} m_2\}]$

2. $[\underline{x}(\alpha), \bar{x}(\alpha)] - [\underline{y}(\alpha), \bar{y}(\alpha)]$

    $= [\min \{\lim_{n\to\infty} m_1 - \lim_{n\to 1} m_2, \lim_{n\to 1} m_1 - \lim_{n\to\infty} m_2\}, \max \{\lim_{n\to\infty} m_1 - \lim_{n\to 1} m_2, \lim_{n\to 1} m_1 - \lim_{n\to\infty} m_2\}]$

3. $[\underline{x}(\alpha), \bar{x}(\alpha)] \times [\underline{y}(\alpha), \bar{y}(\alpha)]$

    $= [\min \{\lim_{n\to\infty} m_1 \times \lim_{n\to\infty} m_2, \lim_{n\to 1} m_1 \times \lim_{n\to 1} m_2\}, \max \{\lim_{n\to\infty} m_1 \times \lim_{n\to\infty} m_2, \lim_{n\to 1} m_1 \times \lim_{n\to 1} m_2\}]$

4. $[\underline{x}(\alpha), \bar{x}(\alpha)] \div [\underline{y}(\alpha), \bar{y}(\alpha)]$

    $= [\min \{\lim_{n\to\infty} m_1 \div \lim_{n\to 1} m_2, \lim_{n\to 1} m_1 \div \lim_{n\to\infty} m_2\}, \max \{\lim_{n\to\infty} m_1 \div \lim_{n\to 1} m_2, \lim_{n\to 1} m_1 \div \lim_{n\to\infty} m_2\}]$

where for any arbitrary interval

$$[\underline{f}(\alpha), \overline{f}(\alpha)] = \left\{ \underline{f}(\alpha) + \frac{\overline{f}(\alpha) - \underline{f}(\alpha)}{n} = m \,\middle|\, \underline{f}(\alpha) \leq m \leq \overline{f}(\alpha), n \in [1, \infty) \right\}.$$

### 3. Moving plate problem

As mentioned earlier we have considered here a parabolic differential equation. Firstly, the said problem is investigated with crisp parameters for the sake of completeness and then the uncertainties are introduced. We have considered a fluid bounded by two parallel plates extended to infinity such that no end effects are encountered. The planar walls and the fluid are initially at rest and we assume that the lower wall is suddenly accelerates in the *x*-direction.

A spatial coordinate system is selected such that the lower wall includes the *xz* plane to which the *y*-axis is perpendicular. The spacing between two plates is denoted by h.

The Navier-Stokes equations for this problem may be expressed as [22]

$$\frac{\partial u}{\partial t} = \nu \frac{\partial^2 u}{\partial y^2} \qquad (1)$$

where, $\nu$ is the kinematic viscosity of the fluid.

Corresponding initial and boundary conditions may be written as below.

<u>Initial condition</u>

$$t=0, \quad \begin{cases} u = U_0, & y = 0 \\ u = 0, & 0 < y \leq h \end{cases}$$

<u>Boundary condition</u>

$$t \geq 0, \quad \begin{cases} u = U_0, & y = 0 \\ u = 0, & y = h \end{cases}$$

The above problem may easily be solved by using well known finite difference method. In the finite difference method, derivatives of the governing equations are usually replaced by difference schemes. First the Forward Time Central Space (FTCS) explicit scheme is considered for the crisp case. As said earlier this is done to have a comparison while analysing the corresponding uncertain problem. Then gradually fuzzy parameters are introduced in the model. Let us choose the time and space step as $\Delta t$ and $\Delta y$ (along *y*-axis) respectively. The velocity term $u_j^n$ has been taken as the second order central difference $\frac{\partial^2 u}{\partial y^2}$, where, *j* and *n* correspond to *y*-axis and time respectively.

Applying FTCS explicit scheme in Eq. (1), we obtain the following discrete expression

$$\frac{u_j^{n+1} - u_j^n}{\Delta t} = \nu \frac{u_{j+1}^n - 2u_j^n + u_{j-1}^n}{(\Delta y)^2} \qquad (2)$$

Eq. (2) can be written as

$$u_j^{n+1} = u_j^n + \nu \frac{\Delta t}{(\Delta y)^2}\{u_{j+1}^n - 2u_j^n + u_{j-1}^n\} \qquad (3)$$

Eq. (3) is the crisp representation of FTCS explicit scheme for Eq. (1) which is the standard well known procedure. In the next section, the main aim of this investigation viz. the uncertain representation of FTCS explicit scheme has been discussed.

### 4. Fuzzy finite difference method

In actual practice, problems depend on experimental observations and operating conditions which are influenced by the systems. In general, the problems contain uncertainty due to the involved parameters, experimental observations and operating conditions of the systems and so these uncertainties may not be avoided. As mentioned earlier that one may handle this uncertainty by probabilistic method but sometime due to the lack of knowledge about the distribution of random values for uncertain parameters one may consider these uncertain parameters as fuzzy/interval form. However it may not be straight forward to handle such uncertainties in term of fuzzy due to complicated fuzzy arithmetic. As such, we target here to handle the problem by taking the uncertain parameters as Triangular Fuzzy Number (TFN) and presented a schematic diagram for the proposed fuzzy finite difference method in Fig. 2.

The schematic diagram for proposed fuzzy finite difference method involves three steps such as input, output and hidden layer. In the input part, uncertain TFN parameters are provided to the system. These parameters are then executed through a step by step process inside the hidden layer which is discussed later in this section. After execution of the hidden layer we get uncertain TFN as output in third step. In Fig. 2, alpha level representation of two fuzzy sets $\tilde{X}$ and $\tilde{Y}$ with their triangular membership functions for fuzzy arithmetic operation has been shown. Deterministic value (crisp) is obtained for $\alpha_4$-level of fuzzy sets whereas for $\alpha_1, \alpha_2$ and $\alpha_3$ level we get the interval values. Output may be generated by considering all possible combinations of alpha levels.

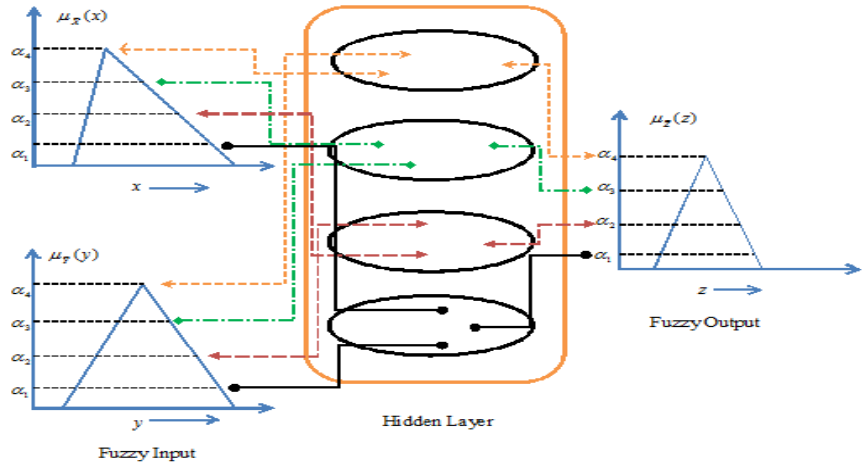

Fig 2. Model diagram of fuzzy finite difference method

Fuzzy numbers are the generalization of interval values for various membership functions $\alpha \in [0, 1]$. Let us assume that these uncertain parameters may be denoted as fuzzy vector $\tilde{x} = (x_1, x_2, \cdots, x_m)^T$, then we may write the fuzzy number $\tilde{x}$ in interval form as $[\underline{x}(\alpha), \bar{x}(\alpha)]$. For every values of $\alpha$ in between 0 and 1 we get an interval and these are represented as $x_\alpha^I$.

Considering these interval representations, the derivatives may be defined as

$$\frac{\partial \tilde{u}}{\partial \tilde{x}} = \frac{\partial [\underline{u}(\alpha), \bar{u}(\alpha)]}{\partial [\underline{x}(\alpha), \bar{x}(\alpha)]} = \frac{\partial u_\alpha^I}{\partial x_\alpha^I}.$$

If we take y-axis as crisp, then the derivative terms may be represented as

$$\frac{\partial \tilde{u}}{\partial y} = \frac{\partial [\underline{u}(\alpha), \bar{u}(\alpha)]}{\partial y} = \frac{\partial u_\alpha^I}{\partial y}$$

$$\frac{\partial^2 \tilde{u}}{\partial y^2} = \frac{\partial^2 u_\alpha^I}{\partial y^2} = \frac{u_{\alpha,j+1}^I - 2u_{\alpha,j}^I + u_{\alpha,j-1}^I}{(\Delta y)^2},$$

$$\frac{\partial \tilde{u}}{\partial t} = \frac{\partial u_\alpha^I}{\partial t} = \frac{u_\alpha^{I,n+1} - u_\alpha^{I,n}}{\Delta t}$$

Using the above representations we may write the fuzzy finite difference scheme for Eq. (1) as

$$u_{\alpha,j}^{I,n+1} = u_{\alpha,j}^{I,n} + v_\alpha^I \frac{\Delta t}{(\Delta y)^2} \left\{ u_{\alpha,j+1}^{I,n} - 2u_{\alpha,j}^{I,n} + u_{\alpha,j-1}^{I,n} \right\} \tag{4}$$

As such, Section 5 gives the stability analysis of the proposed uncertain scheme and the uncertain solution of the said problem has been investigated in section 6.

## 5. Stability Analysis

As fuzzy numbers are the generalization of interval values, so we have discussed here the stability analysis in term of interval values for various membership functions. Accordingly, Eq. (4) may be represented in matrix form as

$$A^I(\Delta t, \Delta y) u_{\alpha,j}^{I,k+1} = B^I(\Delta t, \Delta y) u_{\alpha,j}^{I,k} \tag{5}$$

where $A^I(\Delta t, \Delta y)$ and $B^I(\Delta t, \Delta y)$ are coefficient matrices which depend on $\Delta t$ and $\Delta y$. Here we have assumed that $A^I(\Delta t, \Delta y)$ is non-singular such that we may introduce another matrix

$$C^I(\Delta t, \Delta y) = (A^I)^{-1}(\Delta t, \Delta y) \times B^I(\Delta t, \Delta y) \tag{6}$$

Using Eqs. (5) and (6) we get

$$u_{\alpha,j}^{I,k+1} = C^I(\Delta t, \Delta y) u_{\alpha,j}^{I,k} \tag{7}$$

Let us consider the initial condition as $u_{\alpha,j}^{I,k+1}(0) = U_0$ then Eq. (7) may be written as

$$u_{\alpha,j}^{I,k+1} = C^I(\Delta t, \Delta y) u_{\alpha,j}^{I,k}$$
$$u_{\alpha,j}^{I,k+1}(0) = U_0 \tag{8}$$

We also assume that the error is $\varepsilon^0$ in initial step to investigate the interval stability. Denoting the corresponding uncertain solution by $\tilde{u}_{\alpha,j}^{I,k}$, one may get

$$\tilde{u}_{\alpha,j}^{I,k+1} = C^I(\Delta t, \Delta y) \tilde{u}_{\alpha,j}^{I,k}$$
$$\tilde{u}_{\alpha,j}^{I,k+1}(0) = U_0 + \varepsilon^0 \tag{9}$$

Now subtracting Eq. (8) from Eq. (9) we have

$$\varepsilon^{k+1} = \tilde{u}_{\alpha,j}^{I,k+1} - u_{\alpha,j}^{I,k+1} = C^I(\Delta t, \Delta y)(\tilde{u}_{\alpha,j}^{I,k} - u_{\alpha,j}^{I,k})$$
$$= C^{I,k+1}(\Delta t, \Delta y)\varepsilon^0 \tag{10}$$
$$\varepsilon^0 = \tilde{u}_{\alpha,j}^{I,k+1}(0) - u_{\alpha,j}^{I,k+1}(0)$$

If there exist a constant $K > 0$ for $U_0 \in \Re^N$ then the following inequality satisfies

$$\|\varepsilon^{k+1}\| = \|C^{I,k+1}(\Delta t, \Delta y)\varepsilon^0\| \leq K\|\varepsilon^0\| \tag{11}$$

and then the interval difference scheme is assumed to be stable.

In other words the necessary and sufficient condition of stability for interval difference scheme is

$$\|C^{I,k+1}(\Delta t, \Delta y)\| \leq K \tag{12}$$

## 6. Case Study

We consider the problem discussed in section 3, where the parameters used are given in Table 1. Here both the crisp and fuzzy parameters are taken into account and the model diagram for the present system is shown in Fig. 3.

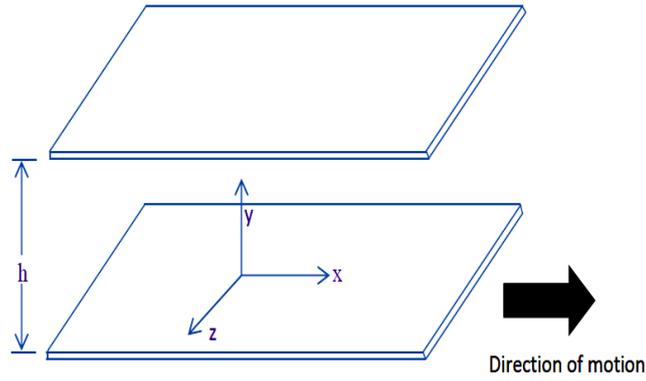

Fig 3. Model diagram of parallel plates system with a suddenly accelerated plane wall

Initially the system is investigated for crisp parameters and then fuzziness has been introduced by using the above discussed proposed method. The fuzziness of the system is studied in detail and the solution set is reported in various cases. Using proposed method the crisp problem is solved for different time (t= 0.18, 0.36, 0.54, 0.72, 0.90 and 1.08) in sec. and the solution sets for the said time are shown graphically in Fig. 4.

Table 1 Crisp and fuzzy parameters for the said system

| Parameters | Crisp [22] | TFN |
| --- | --- | --- |
| $\nu$ | 0.000217 | [0.000180, 0.000217, 0.000250] |
| $h$ | 40 | [30, 40, 50] |
| $U_0$ | 40 | [30, 40, 50] |

Now to study the uncertainty of the system various cases have been discussed by considering different combinations of fuzzy values. The left, centre and right values of the fuzzy solutions are depicted in Figs. 5 to 11.

Different combinations of fuzzy numbers viz. one parameter, two parameters and three parameters are considered as below.

Case 1 (one parameter is fuzzy)

Here only one parameter has been assumed as fuzzy and the obtained results are depicted in Figs. 5 to 7. In Fig. 5, only the kinematic viscosity of the fluid is taken as fuzzy and obtained results for left, centre and right solutions at time, t= 0.18, 0.54, 0.90 sec. are shown. In Fig. 6 height of the plate is considered as fuzzy and the solution set are shown for different time intervals t= 0.18, 0.54, 0.90 sec. It is observed that the uncertain width of the solution set decreases in Fig. 6, as we move from left to right on velocity axis. Whereas, in Fig. 7 initial condition is assumed to be fuzzy and it is seen that the width of the solution set increases as we move from left to right on velocity axis. Again if one move on time t= 0.18, 0.54, 0.90 sec., it is seen that the mean width of the uncertainty increases. Although we get same pattern of curves for three different problems, the uncertain width varies drastically from problem to problem. So the sensitivity of the uncertain parameters changes accordingly.

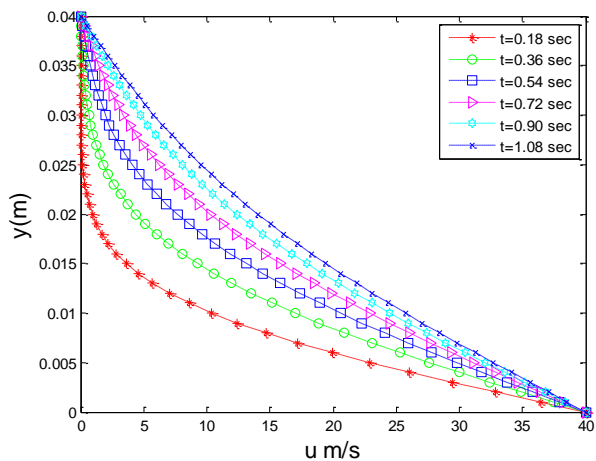
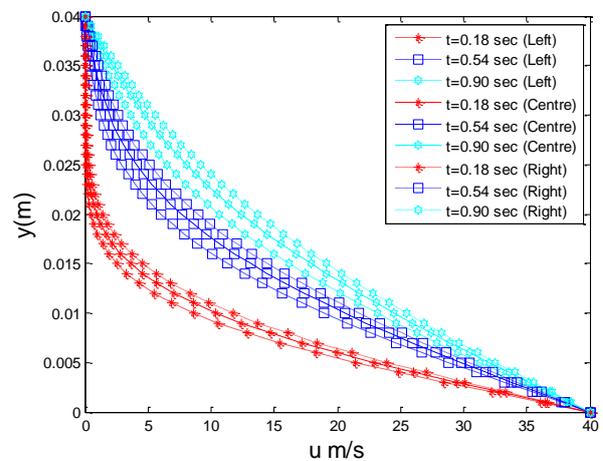

Fig 4. Crisp solution of the system

Fig 5. Solution of the system when only $\nu$ is fuzzy

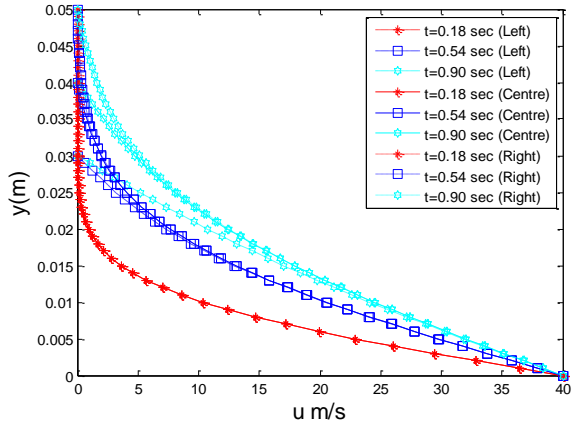 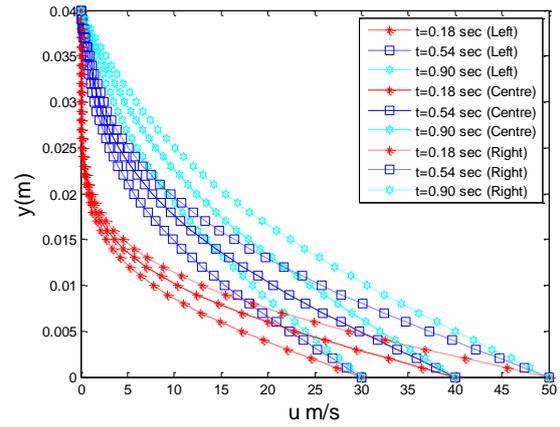

Fig 6. Solution of the system when only $h$ is fuzzy

Fig 7. Solution of the system when only $U_0$ is fuzzy

Case 2 (two parameters are fuzzy)

In this case three problems depending upon two parameters as fuzzy viz. i) kinematic viscosity and height of the parallel plates, ii) height of the parallel pates and initial condition, and iii) kinematic viscosity and initial condition are assumed as fuzzy. Solution sets for different problems are plotted in Figs. 8 to 10. In Fig. 8, kinematic viscosity and height of the plates are taken as fuzzy. It is observed that the width of the uncertain solutions decreases with the velocity axis. In Fig. 9 height of the plates and initial condition are considered as fuzzy and it is seen that the width of the uncertain solutions increases as we move from left to right on velocity axis. Similarly in Fig. 10 kinematic viscosity and initial condition are assumed to be fuzzy. Further it may be noted that the width of the uncertain solutions increases as we move from left to right on velocity axis. If we see the variation of the uncertain solutions along time t= 0.18, 0.54, 0.90 sec., then one may be note that the mean width of the uncertainty also increases for all three problems.

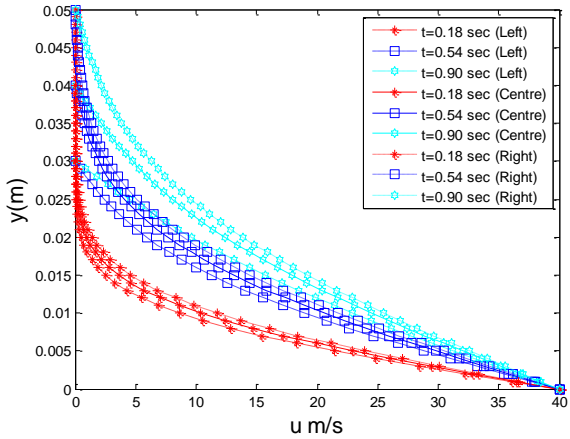

Fig 8. Solution of the system when only $v$ and $h$ are fuzzy

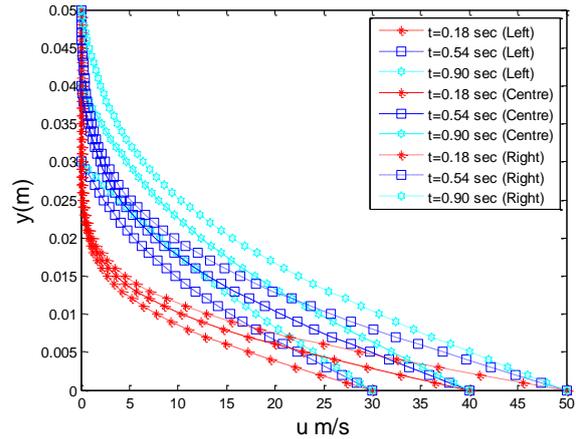

Fig 9. Solution of the system when only $h$ and $U_0$ are fuzzy

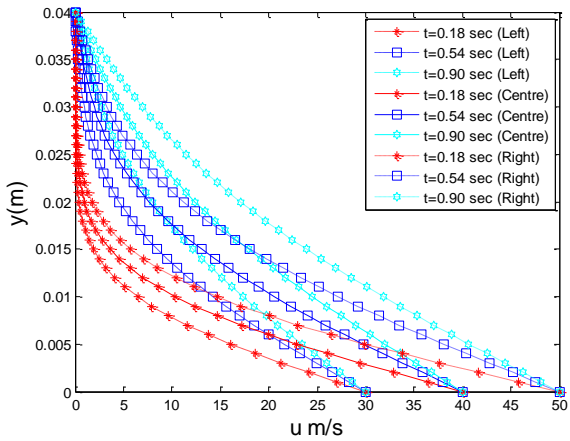

Fig 10. Solution of the system when only $v$ and $U_0$ are fuzzy

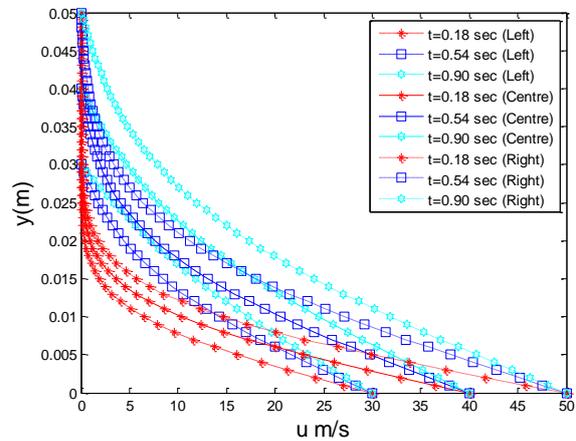

Fig 11. Solution of the system when $v$, $h$ and $U_0$ are fuzzy

#### Case 3 (three parameters are fuzzy)

Finally we consider all three parameters as fuzzy and the obtained set of solutions are depicted in Fig. 11. Here it may be noted that the uncertain mean width of the solution is more when all three parameters (Fig. 11) are fuzzy as compare to the combination of any two parameters (Figs. 8 to 10) as well as the case where one parameter are assumed as fuzzy.

From the above investigation we may observe that the height and initial conditions are more sensitive and it is found that the uncertainty drastically increases with increase in time.

## 7. Conclusion

In this paper, the uncertain parabolic differential equation has been solved numerically in connection with the fuzzy parameters by using proposed fuzzy finite difference method. The concept of fuzziness is hybridised with the well-known finite difference method and the fuzziness of the parameters are managed by using newly developed limit method. For the sake of completeness and validation, stability of the proposed method has also been analysed. Various combinations of uncertain fuzzy parameters are taken into account and uncertain solutions are reported here. It is found that the widths of the uncertainty in solution sets changes drastically when all the parameters are taken as fuzzy.

## 8. Acknowledgement

The authors would like to thank BRNS (*Board of Research in Nuclear Sciences*), Department of Atomic Energy, (DAE), Govt. of India for providing fund to do this work.